\documentstyle[11pt,paspconf,epsf]{article}

\begin{document}

\title{Why Are Halo Density Profiles Stable at Formation?}

\author{Guillermo Gonz\'alez-Casado}
\affil{Departamento  de    Matem\'atica    Aplicada  II, Universidad
Polit\'ecnica de Catalu\~na, Pau  Gargallo 5, 08028 Barcelona, Spain}

\author{Andreu Raig and Eduard Salvador-Sol\'e}

\affil{Departamento de Astronom\'\i a y Meteorolog\'\i a, Universidad de
Barcelona, Av.~Diagonal 647, 08028 Barcelona, Spain}

\begin{abstract}
We analyze the physical justification of the picture proposed by
Salvador-Sol\'e et al.~in these proceedings for the time evolution of
the universal density profile of dark-matter halos. According to this
picture, halos have at formation a stable (i.e.~independent of mass
and time) dimensionless density profile, the characteristic length and
density scales of the profile depending on the underlying
cosmogony. Subsequent evolution is driven by mass accretion onto the
outskirts of halos and can be characterized simply by the increment of
halo radius with time and the corresponding decrease of the critical
density of the universe. We find this picture to be a reasonable good
description of the expected evolution of halos in hierarchical models
of structure formation.
\end{abstract}

\section{Introduction}

Relying on data from high resolution N-body simulations performed by
Navarro, Frenk \& White (1997, NFW), Salvador-Sol\'e et al.~(1998,
S98) derived the laws for the evolution of the dimensionless
characteristic length and density scales ($x_{\rm s}$ and $\delta_{\rm
c}$ respectively) of the universal density profile of dark-matter
halos proposed by NFW. For a halo of mass $M$ at time $t$ (redshift
$z$), those laws have the following expressions:
\begin{eqnarray}
x_{\rm s}(t,M)&=& x_{\rm sf}\,{R(t_{\rm f})\over R(t)}\, ,\label{xs} \\
\delta_{\rm c}(z,M)&=&\delta_{\rm cf}\,{\Omega(z)(1+z_{\rm f})\over 
\Omega(z_{\rm f})(1+z)}\, ,\label{dc}
\end{eqnarray}
where $R(t)$ is the virial halo radius at $t$, $\Omega(z)$ is the
cosmic density at $z$ in units of the critical density of the universe
($\rho_{\rm crit}$), and $t_{\rm f}$ and $z_{\rm f}$ are the
formation time and formation redshift of the halo, respectively. It is
important to remark that equations (\ref{xs}) and (\ref{dc}) are two
independent fitting formulae.  N-body simulations show that $x_{\rm
s}$ and $\delta_{\rm c}$ are, in fact, linked through the condition
that the mean internal density of halos within the virial radius is
equal to about $200\rho_{\rm crit}$ (implying that halo density
profiles are one-parametric functions). Equations (\ref{xs}) and
(\ref{dc}) are found to be consistent with that property for different
cosmological models (cf.~S98).

It is evident that the proportionality constants $x_{\rm sf}$ and
$\delta_{\rm cf}$ (which are cosmogony dependent) correspond to the
values of $x_{\rm s}$ and $\delta_{\rm c}$ when halos form,
respectively. Therefore, the values of the scale radius, $r_{\rm
s}=x_{\rm s}R(t)$, and of the characteristic density, $\rho_{\rm
c}=\delta_{\rm c}\rho_{\rm crit}(z)$, of halo density profiles are set
at the halo formation time. Halo formation is basically characterized
by the last major merger yielding a substantial re-arrangement of the
internal structure of merging halos (see Salvador-Sol\'e et al.~in
these proceedings for a quantitative definition). After formation,
subsequent evolution by matter accretion (secondary infall and/or
minor mergers) does not change the values of $r_{\rm s}$ and
$\rho_{\rm c}$. However, since $\rho_{\rm crit}(z)$ decreases with
time while $R(t)$ increases accordingly, the values of $x_{\rm s}$ and
$\delta_{\rm c}$ change as described by equations (\ref{xs}) and
(\ref{dc}).

In the present work we develop a physical model with the aim of
checking the validity of the picture proposed for the evolution of the
scaling parameters of halo density profiles. On the other hand, we
will try to determine for a set of cosmogonies analyzed the
corresponding shape of the stable density profile of halos at their
formation time. 

\section{The Structure of Halos Formed by Binary Major Mergers}

In this section we describe the model linking the scaling parameters
of the density profile of new formed halos and their progenitors. Let
us assume that halos form essentially through the major merger of two
halo progenitors (this approach is justified in the next section). In
practice, as a major merger we mean that if a halo has a formation
mass $M_0$, then its two halo progenitors have masses $M_1$ and $M_2$
(with $M_1\ge M_2$) so that $M_2/M_1 > \Delta_m = 0.6$. The threshold
$\Delta_m$ between major and minor mergers is an empirical parameter
introduced to allow for a better motivated definition of the formation
time of halos. The value of $\Delta_m=0.6$ has been derived by fitting
the mass-density correlation of halos in high resolution N-body
simulations (see S98).

Consider a halo of mass $M_0$ with internal energy $U_0$ at its
formation time, $t_{\rm f}$, and the system formed by its two halo
progenitors at the time they reach turnaround. The total energy of the
system at turnaround can be written as:
\begin{equation}
E_{\rm ta}=U_1+U_2+E_{12}\, ,
\label{eta}
\end{equation}
where $U_i$ is equal to the sum of the internal energy of the
virialized mass of the $i$-th halo progenitor at turnaround plus the
total energy of the mass that will be accreted onto that halo from
turnaround until $t_{\rm f}$. The last term in the right-hand-side of
equation (\ref{eta}) gives the mutual gravitational interaction plus orbital
kinetic energy of the system at turnaround.

If there is not significant mass-loss neither during the formation
process of $M_0$, nor during mass accretion onto halo progenitors from
turnaround until $t_{\rm f}$, then from energy conservation one has
\begin{equation}
U_0=U_1+U_2+E_{12}\, ,
\label{en}
\end{equation}
and also for $i=1,2$ one can approximate $U_i$ by the internal energy
of halo progenitors at $t_{\rm f}$. Therefore, for $i=0,1,2$ one can write
\begin{equation}
U_i=-{1\over 2}{G M_i^2\over R_i} F(x_{{\rm s}i})\, ,
\label{ui}
\end{equation}
where $M_i$ and $R_i$ are the mass and the radius of each halo at
$t_{\rm f}$, respectively, and $M_1+M_2=M_0$. In writing equation
(\ref{ui}) it has been assumed that halos are virialized systems
described by a universal density profile, which determines the specific
expression of function $F(x_{\rm s})$.

Concerning the expression of $E_{12}$, it can be written as
\begin{equation}
E_{12}=\left[{S\over 2(1-M_2/M_1)}-1\right]{GM_1M_2\over D_{\rm m}}\, ,
\label{e12}
\end{equation}
where $D_{\rm m}$ is the turnaround separation between halo-progenitor
centers. In the well known point-mass approximation and taking $t_{\rm f}$
equal to one orbital period one gets:
\begin{equation}
D_{\rm m}^3=(2-S)^3 G M_0\left({t_{\rm f}\over 2\pi}\right)^2\, .
\label{dm}
\end{equation}
The so-called circularity parameter, $S$, appearing in equations
(\ref{e12}) and (\ref{dm}) takes into account that halo progenitors
can merge following a non-radial orbit. For an elliptical orbit of
eccentricity $e$, one has $S=1-e$. Thus, a radial orbit corresponds to
$S=0$ while a circular orbit to $S=1$. For an object moving in an
arbitrary orbit of total energy $E$ one has $S=J^2/J_c^2(E)$ (Merritt
1985), where $J$ is the angular momentum of the orbit and $J_c(E)$ is
the angular momentum of a circular orbit of energy $E$. Non-radial
mergers are caused by external torques from the large-scale matter
distribution surrounding halo progenitors. In hierarchical structure
formation scenarios one expects that low-mass halos will merge in
non-radial orbits while massive halos will tend to merge in nearly
radial orbits.  On the other hand, the mutual tidal interaction
between merging halos slows-down the orbital motion and the final
merger time will be longer than the value computed for point masses
(eq.~[\ref{dm}]). In practice, this can be accounted for by assuming a
non-null value of $S$, which for a fixed $D_{\rm m}$ increases $t_{\rm
f}$ by a certain amount. The larger the mass of merging halos the
stronger the expected effect of mutual tides on the merger time.
Therefore, the combined action of external torques and mutual tides
between halo progenitors can be described by a value of $S$ different
from zero for a wide mass range. The exact mass dependence of $S$ is
difficult to predict, but in a first approximation the influence of
both external torques and mutual tides on the final results can be
addressed by taking a constant value of $S$ over the whole mass range
analyzed.

Substituting equations (\ref{ui}), (\ref{e12}), and (\ref{dm}) into
equation (\ref{en}) one derives the following equation:
\begin{eqnarray}
F(x_{\rm sf})&=&\left({M_1\over M}\right)^{5\over 3} F(x_{\rm s1}) 
+\left({M_2\over M}\right)^{5\over 3} F(x_{\rm s2}) \nonumber \\ 
&+& (2-S)^{-1}\left[
1-{S\over 2(1-M_2/M_1)}\right]\left({\pi\over 5 \tau}\right)^{2\over 3}
{M_1 M_2\over M^2}\, ,
\label{solve}
\end{eqnarray}
where $\tau=t_{\rm f}H(t_{\rm f})$ and H(t) is the Hubble parameter at
$t$. In the left-hand-side of equation (\ref{solve}) we have written
$x_{\rm sf}$ instead of $x_{{\rm s}0}$ because the latter corresponds
to the value of the dimensionless scale radius of $M_0$ at its
formation time. The expression of $F(x_{\rm s})$ for spherically
symmetric halos in hydrostatic equilibrium with an isotropic velocity
distribution is completely determined by the halo density profile,
$\rho(r)$, which is assumed universal. The different shapes for that
universal profile considered in the present work are particular cases
of the following general expression (Zhao 1996)
\begin{equation}
\rho(r)={\rho_{\rm c} \over
y^{\alpha}(1+y^{\beta})^{{\gamma-\alpha\over\beta}}}\, ,
\label{rho}           
\end{equation}
where $y=r/r_{\rm s}$ and $r$ is the radial distance to the halo center.
From this density profile one can infer the corresponding mass profile
$M(r)$, the one-dimensional velocity dispersion profile $\sigma (r)$, and
finally, one can compute $F(x_{\rm s})$ from the following expression:
\begin{equation}
F(x_{\rm s})=-{U\over G M^2/(2R)}={c\over [m(c)]^2}\left[
\int^c_0 m(y)\hat{\rho}(y)y\,{\rm d} x 
- c^3\hat{\rho}(c)\hat{\sigma}^2(c) \right]\, ,
\label{efe}
\end{equation}
where $\hat\rho(y)=\rho(r)/\rho_{\rm c} $, $m(y)=M(r)/(4\pi r_{\rm
s}^3\rho_c)$, $\hat\sigma(y)=\sigma(r)/\sqrt{4\pi Gr_{\rm s}^2
\rho_c}$, and $c=x_{\rm s}^{-1}$ is the halo concentration parameter.

To sum up, we detail the steps followed by the practical implementation
of the present model:

{\bf Step 1:} Starting from a halo of mass $M$ at a fixed time $t$, we
compute by means of the S98 clustering model the halo formation time
$t_{\rm f}$, its formation mass $M_0$, the typical mass of its halo
progenitors, $M_1$ and $M_2$, and the formation time of the latter,
$t_{\rm f1}$ and $t_{\rm f2}$.

{\bf Step 2:} An empirical value of the dimensionless scale radius
(denoted by $x_{\rm sf}^e$) is assumed for halo progenitors of any mass
at their formation time. This is equivalent to setting a stable
density profile at formation for all progenitors considered if a
universal halo density profile is assumed. 

{\bf Step 3:} The value of $x_{\rm sf}^e$ is evolved from the
formation time of each halo progenitor $t_{{\rm f}i}$ till their
merger time corresponding to the formation time $t_{\rm f}$ of the
resulting new halo $M_0$. This yields the dimensionless scale radii
for the density profile of halo progenitors, $x_{{\rm s}1}$ and
$x_{{\rm s}2}$, at time $t_{\rm f}$ that will be inserted into
equation (\ref{solve}). The evolution of $x_{\rm sf}^e$ can be
computed from either equation (\ref{xs}) or equation (\ref{dc}). We
have checked that both yield consistent results to within typically $10\%$.

{\bf Step 4:} Solving equation (\ref{solve}) for $x_{\rm sf}$ we
derive the theoretical dimensionless scale radius for the density
profile of $M_0$ at formation (denoted by $x_{\rm sf}^t$) which has to
be compared with the corresponding empirical value assigned to halo
progenitors in step 2, $x_{\rm sf}^e$.

\section{Progenitor Masses and Binary Major Mergers}

As stated at the beginning of the previous section, we assume that
major mergers tracing the formation of new halos are essentially binary.
This can be justified in the framework of the clustering model developed
by S98. 

Consider a halo of mass $M_1$ that incorporates a certain amount of
mass $\Delta M$ at time $t$. According to S98, if $\Delta M/M_1 >
\Delta_{\rm m}$ then $M_1$ is destroyed and a new halo of mass
$M_0=M_1+\Delta M$ is formed by a major merger, while if $\Delta M/M_1
\le \Delta_{\rm m}$ the halo $M_0=M_1+\Delta M$ is the result of the
evolution of the halo $M_1$ by mass accretion and/or minor mergers, 
hence $M_1$ is not destroyed, it simply changes its mass to $M_0$.

Now assume that, in the case of major mergers, the mass $\Delta M$
always comes from a single halo that hereafter we denote by $M_2$. It
is easy to see that
\begin{equation}
{\Delta_{\rm m}M_0\over 1+\Delta_{\rm m}} < M_i\ < {M_0\over 1+\Delta_{\rm m}}
\, ,\quad {\rm for\ \ } i=1,2\, .
\end{equation}
This interval limiting the mass of halo progenitors is centered at
$M_0/2$, and $|M_1-M_0/2|=|M_2-M_0/2|$, or equivalently, $M_1$ and
$M_2$ are at the same distance of $M_0/2$. Therefore, the distribution
function of halo progenitor masses should be a symmetric function
around $M_0/2$. This distribution function can be derived from the S98
clustering model and, hence, its symmetry can be checked (see Raig et
al.~1998). We find that indeed the symmetry is fulfilled to within
$3\%$ and this result is independent of redshift and of the cosmogony
considered. The symmetry of the distribution function of halo
progenitors strongly supports that major mergers are essentially
binary.

\section{Results}

In this section we will present the results of the comparison between
the dimensionless scale radius at formation for halos of mass $M$ and
their progenitors ($x_{\rm sf}^t$ and $x_{\rm sf}^e$ respectively),
according to the model described in $\S 2$. This comparison will be
performed for different cosmogonies and a fixed universal density law
of the form given by NFW. Subsequently, we will consider the
implications of assuming a different universal density profile. The
set of cosmogonies analyzed in the present work is described in
columns (1) to (5) of Table 1. We have considered a standard biased
cold dark matter model (SCDM), a flat cold dark matter model with
non-null cosmological constant ($\Lambda$CDM), and a power-law model
with power index $n=-1$ and flat geometry. These models were also
analyzed by NFW and S98. In Table 1, $\Omega_0$ is the matter density
parameter, $\Lambda_0$ is the cosmological constant in units of
$3H_0^2$, where $H_0= 100\, h$ km s$^{-1}$ Mpc$^{-1}$, and $\sigma_8$
is the rms mass fluctuation in spheres of radius $8\, h^{-1}$ Mpc.

\begin{figure}
\plotone{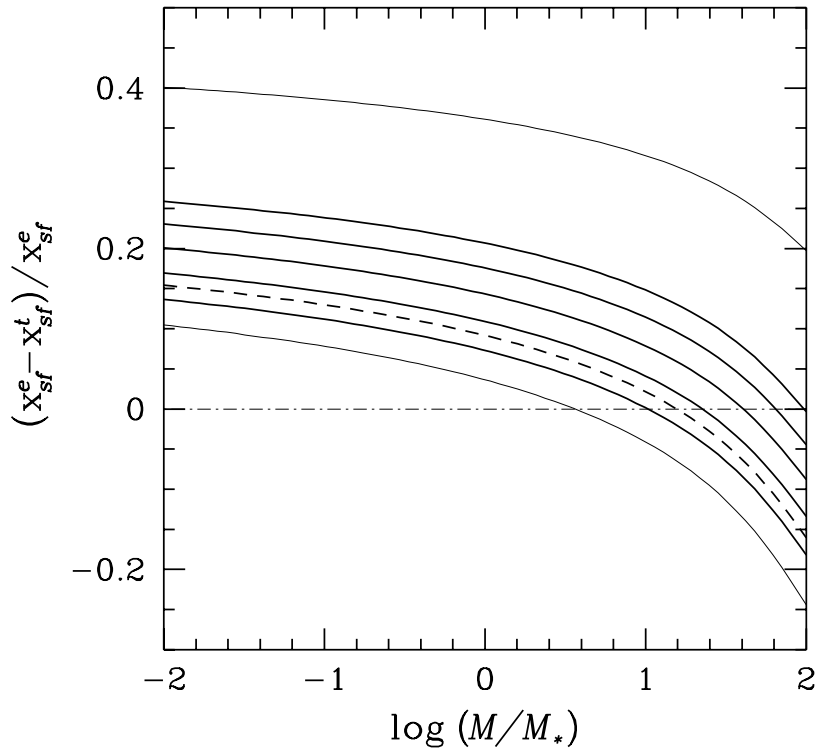}
\caption{Fractional difference between the dimensionless scale radius
at formation for halos of mass $M$ and their progenitors ($x_{\rm
sf}^t$ and $x_{\rm sf}^e$ respectively) in the case of an SCDM
cosmology and assuming an NFW universal density law for dark-matter
halos. The different thick solid lines correspond to different values
of the parameter $S$ ranging from $S=0$ (upper curve) to $S=1$ (lower
curve) in steps of $0.25$. They were derived by taking $x_{\rm sf}^e$
equal to the best fit value obtained in S98 (see Table 1). Thin solid
lines correspond to twice (upper) and half (lower) the latter value
for $S=0$. Finally, the thick dashed line has been derived for the
value of $x_{\rm sf}^e$ that minimizes the maximum fractional
difference in the whole mass range (denoted as $x_{\rm sf}^m$ in the
text). The shape of this latter curve is independent of the value of
$S$ assumed.}
\label{scdm}
\end{figure}

\begin{figure}
\plotone{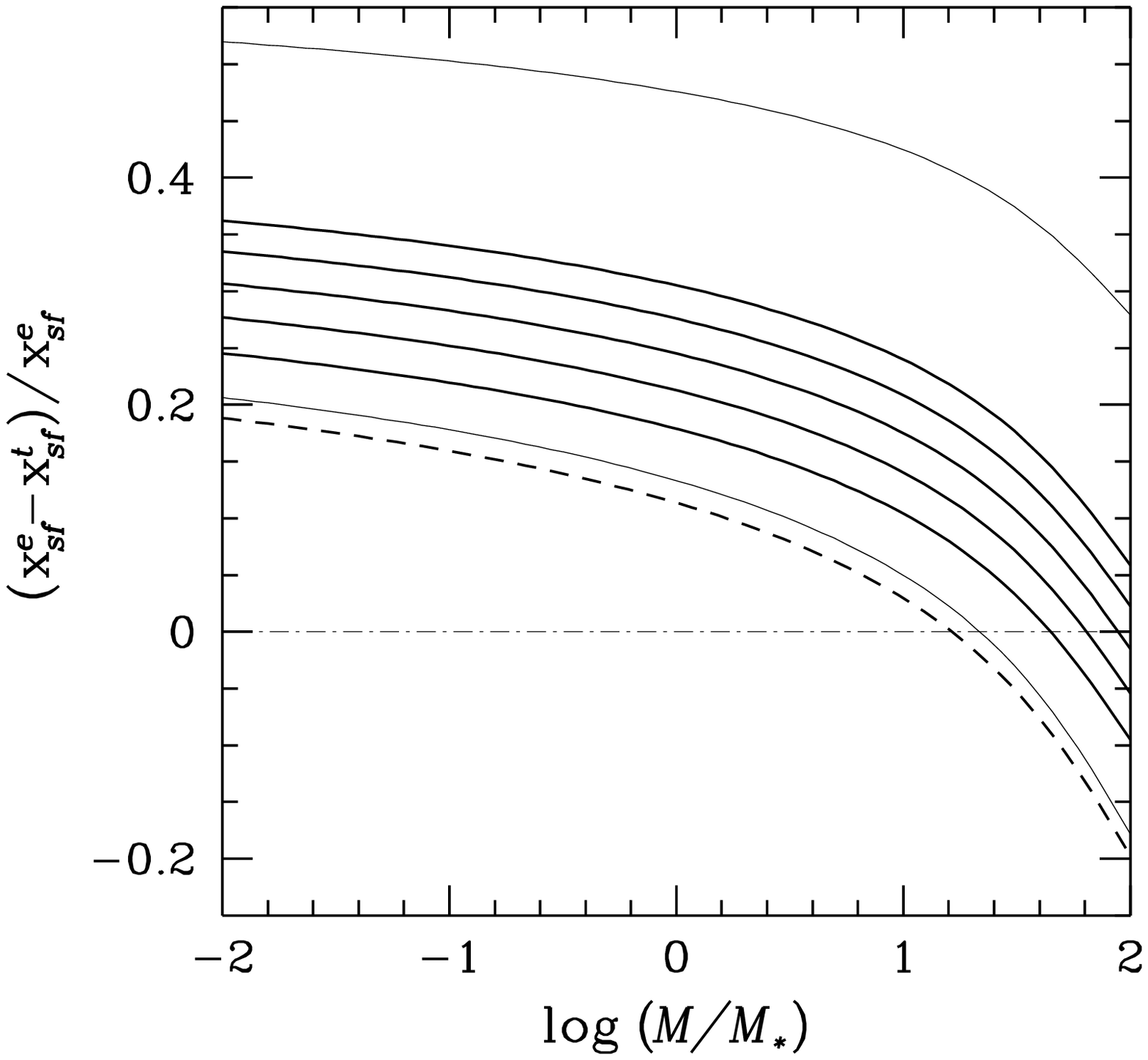}
\caption{Same as Fig.~\ref{scdm} but for the $\Lambda$CDM model.}
\label{lcdm}
\end{figure}

In Figures \ref{scdm} and \ref{lcdm} we show the fractional difference
between $x_{\rm sf}^t$ and $x_{\rm sf}^e$ for the SCDM and the
$\Lambda$CDM models, respectively, assuming a universal NFW density
profile. Similar curves are found for the power-law model. Results are
plotted as a function of halo masses at $z=0$ scaled to the
characteristic mass $M_\ast$ defining a unity rms density
fluctuation. For cosmogonies which are self-similar or close to
self-similar the results expressed in this way are redshift
independent. Thick solid lines correspond to different values of the
parameter $S$ and were derived assuming a value of $x_{\rm sf}^e$
equal to the best fit dimensionless scale radius found by S98 from
their analysis of the NFW N-body simulations. The concrete values are
shown in column (7) of Table 1, and in column (8) are listed the
$90\%$ confidence intervals resulting from the $\chi$-square fit.

As can be seen from Figures \ref{scdm} and \ref{lcdm} the fractional
difference is not far from zero as expected if the dimensionless scale
radius of halos were nearly fixed at formation. It is important to
notice that taking into account the dynamical effects of external
torques and mutual tides ($S>0$) tends to improve the consistency with
a universal value of $x_{\rm sf}$. The maximum fractional difference
is typically between $20\%$ and $35\%$ and is found for low-mass
halos. For massive halos the agreement is quite satisfactory.  On the
other hand, another interesting result is that our model can be used
to predict the value of $x_{\rm sf}^e$ which is in better agreement
with the stability of halo density profiles at formation. As we show
in Figures \ref{scdm} and \ref{lcdm} (upper and lower thin solid
lines), a significant change in the value of $x_{\rm sf}^e$ leads to a
significant variation in the resulting fractional error.
Consequently, we have derived the value of $x_{\rm sf}^e$ that
minimizes the maximum fractional difference in the whole mass range
analyzed. The results, denoted by $x_{\rm sf}^m$, are shown in column
(6) of Table 1 for the case of $S=0.5$, an intermediate value of the
circularity parameter.  The corresponding fractional error curve in
Figures \ref{scdm} and \ref{lcdm} has been plotted as a thick dashed
line. The values of $x_{\rm sf}^m$ are systematically smaller than the
best fit values from S98 but always within the $90\%$ confidence
intervals from Table 1.

\begin{table}
\caption{Scaling of the NFW Density Law for Different Cosmogonies}
\begin{center}
\begin{tabular}{lrrrrccc}
\tableline
$P(k)$ & $\Omega_0$ & $\Lambda_0$ & $\sigma_8$ & $h\,\,$ & $\,x_{\rm sf}^m\,$ & $\,x_{\rm sf}^\ast\,$ & $90\%$ c.i. \\
\ ${}^{(1)}$ & ${}^{(2)}$ & ${}^{(3)}$ & ${}^{(4)}$ & ${}^{(5)}$ & ${}^{(6)}$ & ${}^{(7)}$ & ${}^{(8)}$ \\
\tableline
SCDM & 1.0 & 0.0 & 0.63 & 0.5 & 0.14 & 0.17 & (0.12,0.25) \\
$\Lambda$CDM & 0.25 & 0.75 & 1.3 & 0.75 & 0.18 & 0.29 & (0.18,0.52) \\
$n=-1.0$ & 1.0 & 0.0 & 1.0 & 0.5 & 0.10 & 0.17 & (0.10,0.30) \\
\tableline
\tableline
\end{tabular}
\end{center}
$\qquad\ \qquad {}^\ast\!$ best fit value from S98 
\end{table}

\begin{figure}
\plotone{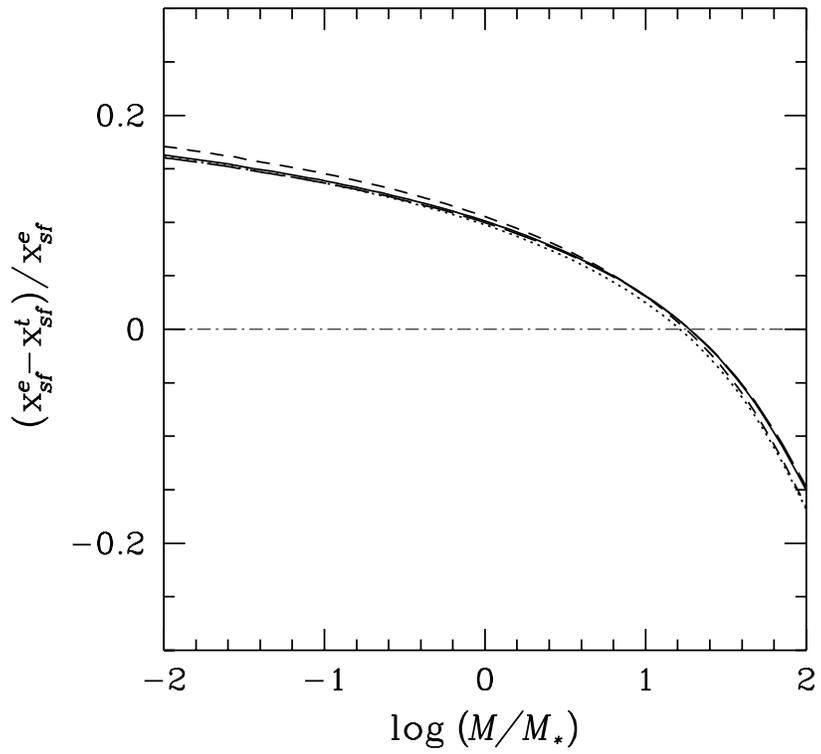}
\caption{The best predictions of our model for the SCDM cosmogony in
the case of the NFW density profile (solid line), the Hernquist
profile (dotted line), the Moore et al.~profile (short-dashed line),
and the 3D King law (long-dashed line).}
\label{perf}
\end{figure}

Recent N-body simulations by Moore et al.~(1997) suggest a profile of
different shape from the one advocated by NFW. This could rise the
question whether our results do depend on the form of the universal
law assumed to describe dark-matter halo density profiles. To clarify
this point we have repeated the previous analysis for different
universal halo density laws in addition to the NFW profile. The latter
corresponds to $(\alpha,\beta,\gamma)=(1,1,3)$ in equation
(\ref{rho}), while the rest of profiles considered are: the Hernquist
(1990) law, $(\alpha,\beta,\gamma)=(1,1,4)$; the three-dimensional
(3D) King law, $(\alpha,\beta,\gamma)=(0,2,3)$; and the profile
proposed by Moore et al. (1997) to fit their high resolution N-body
simulations, $(\alpha,\beta,\gamma)=(1.4,1.4,2.8)$. One can scale the
values of $x_{\rm s}$ and $\delta_{\rm c}$ from the NFW profile to the
rest of density laws by assuming that the maximum in the different
circular velocity profiles is at the same physical distance to halo
center. The Hernquist and NFW profiles are similar in the innermost
regions of halos. The Hernquist profile has a different slope from the
NFW profile in the outermost regions of halos although it was found to
provide a satisfactory fit to halo density profiles (NFW).  On the
other hand, in the outermost regions the NFW and the 3D King laws are
similar.  The 3D King law has a flat core which up to now has not been
found in any cosmological N-body simulation (within the nominal
resolution scale), but this profile provides a good description of the
distribution of galaxies in clusters. For the present study, the King
law can be considered as providing a limiting case for the slope of
the density profile in the central region of halos. Since the exact
value of this slope is still a subject of debate, we have included in
our study the profile suggested by Moore et al. (1997).

For each density profile we have derived the value of the
dimensionless scale radius at formation that, according to our model,
minimizes the maximum fractional difference between $x_{\rm sf}^e$ and
$x_{\rm sf}^t$ in the mass range considered.  As illustrated by Figure
\ref{perf} corresponding to the SCDM cosmology, all density profiles
lead to essentially the same result concerning the fractional
difference. On the other hand, with respect to the value of $x_{\rm
sf}^m$ inferred for the NFW profile, the values obtained for the
other density laws analyzed differ in less than $20\%$.

\section{Conclusions}

The results of our study confirm the picture proposed in S98 and by
Salvador-Sol\'e et al. in these proceeding to describe the scaling
evolution of universal halo density profiles. Halos form with a
universal value of the dimensionless scale radius and scale density,
$x_{\rm sf}$ and $\delta_{\rm cf}$, i.e., halos have stable density
profiles at formation. Subsequently, halos evolve by accretion and the
corresponding evolution of $x_{\rm s}$ and $\delta_{\rm c}$ is well
described by equations (\ref{xs}) and (\ref{dc}). Our study has
covered a mass range of four decades, typically from the mass of
galaxies to the mass of galaxy clusters. We have shown that the
validity of the proposed picture is not affected by the specific form
assumed for the universal density law of dark-matter halos. For
density laws of the form (\ref{rho}) we always reach at essentially
the same best model-predicted fractional difference between the values
of $x_{\rm sf}$ for halo progenitors and the new halos they form. We
have derived for different cosmogonies the values of $x_{\rm sf}$
which lead to the best consistency with the previous picture. The
resulting values are found to be in agreement with the best fit ones
obtained by S98 from the analysis of N-body simulations. 

According to the previous conclusions, in order to characterize the
scaling evolution of halo density profiles, a first physical principle
such as energy conservation during halo formation has to be taken into
account. But at the same time, one also requires a clustering model
properly describing halo formation and evolution through an adequate
distinction between major mergers, tracing the formation of new
objects, and minor mergers, corresponding to accretion onto otherwise
relaxed halos. On the other hand, the exact form of the universal
density profile of halos, which can not be determined by our analysis,
would be set by more complicated processes than those included in the
present study. Those processes would probably be linked to violent
relaxation during halo formation in the framework of approximately
self-similar cosmogonies.

\acknowledgments

This work has been supported by the Direcci\'on General de
Investigaci\'on Cient\'\i fica y T\'ecnica under contract PB96-0173
and by the Universidad Polit\'ecnica de Catalu\~na research project
PR9707.

\end{document}